\documentclass{article}

\begin{document}

\def\giorno{11/2/2002}

\renewcommand{\theequation}{\arabic{section}.\arabic{equation}}
\newenvironment{theorframe}[1]{\refstepcounter{equation}\par\medskip\noindent%
   {\bf #1~~\theequation}\hspace{5pt}\em}%
   {\em\par\medskip}

\newenvironment{lemma}{\begin{theorframe}{Lemma}}{\end{theorframe}}
\newenvironment{proposition}{\begin{theorframe}{Proposition}}{\end{theorframe}}
\newenvironment{theorem}{\begin{theorframe}{Theorem}}{\end{theorframe}}
\newenvironment{corollary}{\begin{theorframe}{Corollary}}{\end{theorframe}}

\def\norma#1{\left\Vert #1\right\Vert}
\def\V{{\cal V}}
\def\U{{\cal U}}
\def\bp{{\bf p}}
\def\bq{{\bf q}}
\def\Fb{{\bf F}}
\def\F{{\cal F}}
\def\L{{\cal L}} 
\def\M{{\cal M}} 
\def\R{{\bf R}}
\def\cb{{\bf c}}
\def\a{\alpha}
\def\b{\beta}
\def\eps{\varepsilon}
\def\ra{{\bf Z}}
\def\de{\delta}
\def\om{\omega}
\def\Om{\Omega}
\def\epsilon{\varepsilon}
\def\La{\Lambda}
\def\sse{\subseteq}
\def\ss{\subset} 
\def\s{\sigma}
\def\S{\Sigma} 
\def\ga{\gamma}
\def\toro{{\bf T}}
\def\T{{\rm T}}
\def\de{\delta}
\def\d{{\rm d}}
\def\phi{\varphi}
\def\pa{\partial}
\def\w{\wedge}
\def\({\left(}
\def\){\right)}
\def\[{\left[}
\def\]{\right]}
\def\^#1{{\widehat #1}}
\def\~#1{{\widetilde #1}}
\def\diff#1#2{\frac{\partial#1}{\partial#2}}
\def\k{\kappa}

\def\sp{\medskip \par\noindent}
\def\EOP{$\triangle$ \par \medskip}
\def\EOR{$\odot$ \par \medskip}
\def\C{{\bf C}}
\def\interno{\hskip 2pt \vbox{\hbox{\vbox to .18
truecm{\vfill\hbox to .25 truecm
{\hfill\hfill}\vfill}\vrule}\hrule}\hskip 2 pt}
\def\Re{{\bf R}}

\def\mapright#1{\smash{\mathop{\longrightarrow}\limits^{#1}}}
\def\mapdown#1{\Big\downarrow\rlap{$\vcenter{\hbox{$\scriptstyle#1$}}$}}
\def\mapleft#1{\smash{\mathop{\longleftarrow}\limits^{#1}}}
\def\mapup#1{\Big\uparrow\rlap{$\vcenter{\hbox{$\scriptstyle#1$}}$}}

%%%%%%%%%%%%%%%%%%%%%%%%%%%%%%%%%%%%%%%%%%%%%%%%%%%%%%%%%%%%%%%%%%%%%%%%%%%

\title{On persistence of invariant tori \\ and a
theorem by Nekhoroshev}

\author{Dario Bambusi\footnote{E-mail: bambusi@mat.unimi.it} \ and \
Giuseppe Gaeta\footnote{Supported by ``Fondazione CARIPLO per la
Ricerca Scientifica'' under project ``Teoria delle perturbazioni per
sistemi con simmetria''. E-mail: gaeta@berlioz.mat.unimi.it} \\ {\it
Dipartimento di Matematica, Universit\`a di Milano} \\ {\it via
Saldini 50, I--20133 Milano (Italy)} }

\date{~}

\maketitle

\noindent
{\bf Summary.} {We give a proof of a theorem by N.N. Nekhoroshev
concerning Hamiltonian systems with $n$ degrees of freedom and $s$
integrals of motion in involution, where $1 \le s \le n$.  Such a
theorem ensures persistence of $s$-dimensional invariant tori under
suitable nondegeneracy conditions generalizing Poincar\'e's condition
on the Floquet multipliers.}

\section*{Introduction.}

A Hamiltonian system in $n$ degrees of freedom having $n$ independent
integrals of motion in involution is integrable \cite{Arn1}.  When the
system has a number $s$ of independent integrals of motion greater
than one, but smaller than the number of degrees of freedom, i.e. $1 <
s <n$, we say the system is ``partially integrable''.

In the early nineties Nekhoroshev stated, under suitable nondegeneracy
conditions, an interesting result on the existence of $s$-parameter 
families of tori in partially integrable Hamiltonian systems with $n$
degrees of freedom ($1<s<n$) \cite{Nek}; this constitutes a bridge between
the Poincar\'e-Lyapunov theorem (case of periodic trajectories, $s=1$) 
and the Liouville-Arnold one (complete integrability, $s=n$).
 
Unfortunately Nekhoroshev never published a proof of this theorem.
When we became interested in the problem, we decided to try to
reconstruct the proof, following a line of thought which is naturally
suggested by Nekhoroshev's formulation of the nondegeneracy
condition. However, after doing this we realized that one can also
reformulate the nondegeneracy condition of the theorem in terms of
standard objects (Floquet multipliers), and obtain a simpler proof:
this is given in the present paper.

\medskip

We recall that, roughly speaking (the precise statement will be
recalled in sect. 1), Nekhoroshev's theorem states that if one has $s$
integrals of motion in involution and an invariant torus $\La \simeq
\toro^s$, then (under suitable nondegeneracy conditions) there is a
$2s$--dimensional symplectic submanifold $N$, with $\Lambda\ss N $,
which is fibered by invariant tori $\Lambda_\b \simeq \toro^s$;
moreover it is possible to build action angle coordinates on
$N$. Finally the invariant tori $\Lambda_\beta$ persist under small
perturbations of the Hamiltonian and of the integrals of motion. The
main problem lies in identifying the relevant nondegeneracy
condition. We find that it can be expressed in terms of Floquet
multipliers of periodic orbits of suitable Hamiltonian vector
fields. In the case of reducible tori the condition can be
reformulated in a purely algebraic form.

\medskip

The {\bf plan of the paper} is as follows. In section 1 we recall the
precise statement of Nekhoroshev's theorem. Section 2 is devoted to a
detailed proof of the main theorem, based on a lemma which provides
``good" coordinates. In section 3 the theory is applied to the case of
{\it reducible tori} (whose definition is recalled there); in such a
case the nondegeneracy condition reduces to a condition on
determinants of $s$-dimensional matrices, see eq.(\ref{simple}).

\subsection*{Acknowledgements}

We thank N.N. Nekhoroshev for discussions on his theorem and for
the encouragement to publish a complete proof of his result.  The work
of GG was supported by {\it ``Fondazione CARIPLO per la Ricerca
Scientifica''} under the project {\it ``Teoria delle perturbazioni per
sistemi con simmetria''}.

\section{Statement of results}

Let $(M,\Om )$ be a symplectic manifold (with symplectic form $\Om$)
of dimension ${\rm dim} (M) = 2n$, differentiable of class $C^r$ ($r
\ge 2$). Let $\Fb_\epsilon := \{ F^\epsilon_1 , ... , F^\epsilon_s \}$
be $s$ real functions on $M$ depending in a $C^r$ way on a small
parameter $\epsilon\in E$, with $E:=(-\epsilon_0,\epsilon_0)$, and
$\epsilon_0>0$. We assume also that the functions $F_i^\epsilon$ are
differentiable of class $C^r$, that they are independent and in
involution, namely that $\left\{F_i^\epsilon,F_j^\epsilon\right\}=0$,
$i,j \in \{ 1,...,s \}$, $\forall\epsilon\in E$.
 
We can then consider the Hamiltonian vector fields $X^\epsilon_i$
generated by the functions $F^\epsilon_i$, namely defined by
$X^\epsilon_i \interno \Om = \d F^\epsilon_i$.

We will {\it assume} that there is a compact and connected 
manifold $\Lambda \ss M$ of dimension $s$, differentiable of class
$C^r$, which is invariant under all the unperturbed vector fields
$X_i:=X_i^0\equiv X_i^\epsilon\big|_{\epsilon=0}$, and such that the $X_i$
are linearly independent at all points $m \in \Lambda$. Obviously this
$\Lambda$ will be a submanifold of $\Fb_0^{-1} (\beta_0)$ for some
$\beta_0 \in \Re^s$.

Note that the condition of independence of the $X_i$ (and of the
$F_i^\epsilon$) on $\Lambda$ is equivalent to the requirement that
$$ 
\eta^\epsilon := \d F^\epsilon_1 \w ... \w \d F^\epsilon_s \not= 0 \ \
\ {\rm on} \ \Lambda \ \ \ {\rm and\ for} \ \ \epsilon=0; 
$$ 
since the $F^\epsilon_i$ are smooth (i.e. $C^1$) functions, the
relation $\eta^\epsilon \not= 0$ holds in a neighbourhood of $\La$ and
for $\epsilon$ small enough.
 
It is actually immediate to see that, being $s$-dimensional,
connected, compact and invariant under the $s$ commuting vector fields
$X_i$, the $C^r$ manifold $\La$ is necessarily a torus: $\La =
\toro^s$; see e.g. the proof of the Liouville-Arnold theorem in
\cite{Arn1}.  Thus, under our assumptions the system has an invariant
$s$-torus.

It is well known that for $s=1$ (i.e. $\La \approx S^1$ is a periodic
orbit), if the Floquet multiplier 1 has multiplicity two, then $\La$
is part of a one-parameter local family of such orbits which persists
under perturbation (Poincar\'e-Lyapunov theorem).  We want to show
that a similar property holds for arbitrary $s$, $1 \le s \le n$; the
main problem lies in identifying the appropriate nondegeneracy
condition.
 
\bigskip

In order to state the theorem we need to fix some notation. 

First of all, using the same notation as in \cite{Arn1}, we will
denote by $g_{\epsilon,i}^t(.)$ the flow of the vector field
$X^\epsilon_i$, and, for $\tau\equiv(\tau_1,\tau_2,...,\tau_s)\in
{\cal T} \sse \Re^s$, we define
\begin{equation}
\label{flu}
g_\epsilon^{\tau}(p)=g_{\epsilon,1}^{\tau_1}\circ
g_{\epsilon,2}^{\tau_2}\circ...\circ g_{\epsilon,s}^{\tau_s}(p) \
. 
\end{equation}
We will also denote 
$$
g^{\tau}:=g^\tau_0\ .
$$

Consider the first homotopy group $\pi_1(\Lambda) = {\bf Z}^s $ of
$\La$, and fix $\alpha = (\a_1 , ... , \a_s ) \in \pi_1(\La)$; then
there exist real constants $(c_1,...,c_s)\equiv\cb$ such that the flow
of the vector field
\begin{equation}
\label{N.1}
X_{\alpha} \ := \ \sum_{i=1}^s \ c_i \, X_i 
\end{equation}
with initial data on $\La$ is periodic of period one (if $\a
\not\equiv 0$) and has closed
trajectories in the homotopy class $\alpha$. Such periodic orbits are
given by $g^{t\cb}(m)$, $t\in[0,1]$, $m\in\Lambda$. We will denote by
$X_\alpha^\epsilon$ the vector field $X_{\alpha}^\epsilon \ := \
\sum_{i=1}^s \ c_i \, X^\epsilon_i $.  

We recall that for a fixed $m \in \La$, the Floquet multipliers of the
periodic orbit $g^{t\cb}(m)$ are the eigenvalues of the Jacobian
matrix $[dg^{\cb}](m)$ (where the differentiation is with respect to
the variable $m$).  It turns out (as discussed in lemma \ref{flo}
below) that the Floquet multipliers of the orbit $g^{t\cb}(m) $ do not
depend on the point $m \in \Lambda$, but only on the vector field
$X_\alpha$. Moreover, due to the symmetries of the problem (see the
proof of lemma \ref{impl} below), $1$ is a Floquet multiplier of such
a periodic orbit with multiplicity at least $2s$.

\begin{theorem}
\label{pert} ({\bf Nekhoroshev}) Let $(M,\Om )$ be a
$2n$-dimensional symplectic manifold. Let $\Fb_\eps := \{ F_1^\eps ,
... , F_s^\eps \}$ be $s$ real functions on $M \times E$,
differentiable of class $C^r$ jointly in  the $M$ variables and in
$\eps$; and denote by $X_i^\eps$ the corresponding Hamiltonian vector
fields.

Assume that: 

\noindent {\rm (i)} there exists a $C^r$ compact and connected
$s$-dimensional manifold $\La \ss M$ invariant under all the
$X_i\equiv X_i^{\eps}\big\vert_{\eps=0}$;

\noindent {\rm (ii)} the functions $F_i^\eps$ are independent and in
involution -- namely $\eta^\eps \not= 0$ and
$\left\{F_i^\eps , F_j^\eps \right\}=0$ $\forall i,j\in \{ 1,...,s
\}$ --  in a neighbourhood $W$ of $\Lambda$, for all $\eps \in E$;

\noindent {\rm (iii)} there exists $\alpha\in\pi_1(\Lambda)$ such that
the Floquet multiplier 1 of the periodic orbits of $X^0_\alpha$ on
$\Lambda$ has multiplicity $2s$.

Then there exists $\eps_*>0$, such that, for all $\eps\in
E_0$, $E_0:=(-\eps_*,\eps_*)$, the following holds true:

\noindent {\rm (1)} In a neighbourhood $U \sse W$ of $\La$ in $M$,
there is a family of symplectic submanifolds $N_\eps \ss U$, of
dimension $2s$, which are fibered over a domain $B \ss \R^s$ having as
fibers $C^r$-differentiable tori $\La_\b^\eps = N_\eps \cap \Fb_\eps^{-1}
(\b) \simeq \toro^s$ (with $\b \in B$). For each $\eps \in E_0$, the
tori $\La_\b^\eps$ are $C^r$-diffeomorphic to $\La$ and invariant
under the $X^\eps_i$; the tori $\La_\b^\eps$ and the manifolds $N_\eps$
depend in a $C^r$ way jointly on
$(\eps,\b) \in E_0 \times B$.

\noindent {\rm (2)} There exist symplectic action angle coordinates
$(I_1^\eps,...,I_s^\eps ; \phi_1^\eps , ... , \phi_s^\eps )$ in $N$,
such that $F_i^\eps \big\vert_{N_\eps} = F_i^\eps \big\vert_{N_\eps}
(I_1^\eps ,..., I_s^\eps )$, with $i=1,...,s$. The coordinates
$(I^\eps ; \phi^\eps )$ depend in a $C^r$ way on $\eps \in E$, and so
do the functions $F_i^\eps \big\vert_{N_\eps}$.

\end{theorem}

\bigskip

\sp {\bf Remark 1.} The theorem applies in particular to the case of a
hamiltonian system with a compact symmetry group acting properly and
freely, and which admits an $Ad^*$ equivariant momentum map. To fix
ideas suppose that $F_1$ is the Hamiltonian of the system and
$F_2,...,F_s$ are the integrals in involution related to the momentum
map. In this case the invariant torus $\Lambda$ is just a relative
periodic orbit (namely a periodic orbit of the system obtained by the
Marsden--Weinstein reduction procedure). Thus in this case the
statement (1) is very close to the Poincar\'e continuation theorem for
relative periodic orbits. This situation has been studied extensively
(see \cite{M,WR} and references therein).  \EOR

\sp {\bf Remark 2.} In the above situation,
if the symmetry group acts linearly, the solutions of the
equations of motion of the Hamiltonian vector fields of $F_2,...,F_s$
are automatically periodic also for initial data outside
$\Lambda$. We stress that this is not assumed here. \EOR

\sp{\bf Remark 3.} A related result, based on topological arguments, has
been recently proved by J.P.Ortega \cite{Ort}, who extended the 
Weinstein--Moser theorem in order to find a lower bound to the number
of relative periodic orbits (i.e. invariant tori of the original system)
close to relative equilibria. \EOR

\sp {\bf Remark 4.} In the same situation one could try to apply KAM
theory for lower dimensional tori \cite{Bou,Eli,Kuk,Poe} (note that
this would require a stronger nondegeneracy condition), and this would
ensure persistence of a Cantor family of invariant tori; on the other
hand we ensure here existence of a {\it continuous} family of
invariant tori. In particular we ensure also persistence of the {\it
resonant} tori. Obviously this is due to the fact that the systems
admit some integrals of motions independent of the Hamiltonian, and
thus our situation is exceptional. \EOR

\sp {\bf Remark 5.} When studying infinite dimensional systems, one meets
cases where a continuous spectrum arises, and this makes KAM theory
non applicable at all. On the other hand theorem \ref{pert} extends
immediately to some infinite dimensional situations of this kind; in
particular it is used in \cite{BV} to construct quasiperiodic
breathers in infinite lattices. \EOR

Our formulation differs slightly from the original one \cite{Nek} as
it is focused to the perturbative frame. In this case the statement
(2) is particularly useful in that it allows to characterize the
dynamics on the invariant tori of the perturbed system. In particular
one has the following

\begin{corollary}
\label{qp}
In the same hypotheses of theorem \ref{pert}, assume also that for some
$\k\in\left\{1,...,s\right\}$, and for some $m\in\Lambda$ one has 
\begin{equation}
\label{nd1}
{\rm det}\left(\frac{\partial^2 F^0_{\k}\big|_{\Lambda}}{\partial
I_j\partial 
I_k}\right)(m) \not=0\ , 
\end{equation}
and fix $\eps$ small enough; then there exists a neighbourhood $B_0$
of $\beta_0$ in $\Re^s$ such that for almost all $\beta\in B_0$ the
flow of $X_{\k}^{\eps}$ on $\Lambda_{\beta}^\eps$ is
quasiperiodic with $s$ frequencies independent over the rationals.
\end{corollary}

\sp {\bf Proof.} By smooth dependence of action angle variables and of
$F^\eps_{\k}$ on $\eps$, eq.\ref{nd1} holds also for $\eps$ small 
but different from zero. It follows that the map from the actions to 
the frequencies is a local isomorphism. \EOP

\section{Proofs}

In this section we will always {\it assume that all the hypotheses of
theorem \ref{pert} hold}, without stating this explicitly in each
lemma. 

\bigskip

In order to prove the theorem stated in the previous section, the
main point is to introduce suitable coordinates in a neighbourhood $U$
of an arbitrary point of $\Lambda$. In order to do that we introduce,
for any $m$ in $\Lambda$ a manifold $\Sigma_m$ of codimension $s$
passing through $m$. By the tubular neighbourhood theorem \cite{Lang},
the manifolds $\Sigma_m$ can be chosen in such a way to define a
foliation in $U$.

\begin{lemma}
\label{coo}
Let $m$ be an arbitrary point in $\La$. There exists a neighbourhood
$\V_m \subset W \subset M$ of $m$ a positive $\epsilon_0$ and, for any
$\epsilon$ with $|\epsilon|<\epsilon_0$ a coordinate map 
$$
\V_m\ni
p\mapsto (\beta,\tau,y)\in\R^s\times\R^s\times\R^{2(n-s)}
$$ 
with the following properties (where we identify a point with its
coordinates):

(i) $F^\epsilon_i (\beta,\tau,y)=\beta_i $, 

(ii) when $\epsilon=0$ the coordinates of $m$ are $(\beta_0,0,0)$

(iii) $(\partial / \partial\tau_i ) \, = \, X^\epsilon_i$, $\forall i=1,...,s$

(iv) $p\in\Sigma_m\ \iff\ p=(\beta,0,y)$

(v) the coordinate map depends smoothly on $m$ and $\epsilon$.
\end{lemma}

\sp {\bf Proof.} First one has that $\Fb_\epsilon$ restricted to $\Sigma_m$ is
a submersion and therefore one can introduce coordinates $(\beta,y) $
in $\Sigma_m$ defined in an open set of $ \R^s\times\R^{2(n-s)}$ with
$\Fb_\epsilon(\beta,y)=\beta$ and such that, when $\epsilon=0$, the coordinates
of $m\in\Sigma_m$ are $(\beta_0,0)$.

For $(\tau , z) $ in some subset of $\R^s\times \Sigma_m$, consider
the map $(\tau,z)\mapsto g_\epsilon^{\tau}(z)$.  Using the coordinates
$(\beta,y)$ in $\Sigma_m$ this can be expressed as a map $(\tau,
\beta, y)\mapsto g_\epsilon^{\tau}(\beta,y)$. It is easily verified that, for
$\epsilon=0$, its differential at $m$ is an isomorphism and therefore,
by the implicit function theorem, the $(\beta,\tau,y)$ are a local
coordinate system at $m$ for any $\epsilon$ small enough. Moreover by
construction they have the properties stated in the lemma. \EOP

The coordinates constructed in this lemma will be called
{\it adapted coordinates based at the point $m$} 

Let us now fix a nontrivial homotopy class $\alpha\in\pi_1(\Lambda)$
and, as in the previous section, consider a vector field $X_\a =
\sum_i c_i X_i$ such that its trajectories are periodic of period 1 on
$\Lambda$ and belong to the homotopy class $\a$.  For ease of writing,
we denote by $\Phi_\epsilon$ the time 1 flow of $X^\epsilon_\alpha$,
i.e. $\Phi_\epsilon (p) = g_\epsilon^{\cb}(p)$ where
$\cb:=(c_1,...,c_s)$. 

Consider now a point $m \in \La$ and introduce adapted coordinates
based at $m$; we write
$(\^\beta_\epsilon,\^\tau_\epsilon,\^y_\epsilon):=\Phi_\epsilon(\beta,\tau,y)$.
Remark that since $\Phi_0(\beta_0,0,0)=(\beta_0,0,0)\equiv m$, by
smooth dependence of solutions on initial data and parameters there
exists a neighbourhood ${\widetilde \V_m}$ of $m$ which is mapped
under $\Phi_\epsilon$ in the domain $\V_m$ of definition of the above
coordinates. We restrict $\Phi_\epsilon$ to such a neighbourhood.

\begin{lemma}
\label{str}
In the adapted coordinates, the map $\Phi_\epsilon: (\b , \tau , y) \to
(\^\beta_\epsilon,\^\tau_\epsilon,\^y_\epsilon) $ is described by
$$ 
\^\beta_\epsilon (\beta,\tau,y) \, = \, \beta \ \ ; \ \ 
\^\tau_\epsilon (\beta,\tau,y) \, = \, \tau \, + \, \bar\tau_\epsilon(\beta,y) \ , 
$$
with $\bar\tau_\epsilon$ a suitable function.
Moreover, 
$$ 
\partial \^y_\epsilon / \partial \tau \ = \ 0 \ . 
$$
\end{lemma}

\sp{\bf Proof.} The first equality is a trivial consequence of the way
the coordinates are defined. To prove the other two equalities fix
$\mu\in\Re^s$ small and 
consider 
$$ 
\^\beta_\epsilon (\beta,\tau+\mu,y)\ \ ; \ \ 
\^\tau_\epsilon (\beta,\tau+\mu,y) \, \ ; \ \ \^y_\epsilon(\beta,\tau+\mu,y)
$$
by definition these are the coordinates of the point
$$
\Phi_\epsilon(\beta,\tau+\mu,y)=\Phi_\epsilon\left(g_\epsilon^{\mu}(\beta,
\tau,y)\right)=
g_\epsilon^{\mu}
\left(\Phi_\epsilon\left(\beta,\tau,y)\right)\right)
= (\beta,\hat\tau_\epsilon(\beta,\tau,y)+\mu,\^y_\epsilon(\beta,\tau,y))\ .
$$ From this
$$
\^\tau_\epsilon(\beta,\tau+\mu,y)=\^\tau_\epsilon (\beta,\tau,y)+\mu\
, \quad \^
y_\epsilon(\beta,\tau+\mu, y)=\^ y_\epsilon(\beta,\tau,y)
$$
which shows that $\^ y_\epsilon$ is independent of $\tau$. To conclude
the proof just put $\bar\tau_\epsilon(\beta,y):=\hat \tau_\epsilon
(\beta,0,y)$. \EOP

It is useful to remark that the Floquet multipliers of the
periodic orbits of $X_{\alpha}$ do not depend
on the initial point of the orbit in $\Lambda$.

\begin{lemma}
\label{flo}
Let $m$ and $m_1$ be two points of $\Lambda$; then the Floquet
multipliers of $g^{t\cb}(m)$ and of $g^{t\cb}(m_1)$ coincide.
\end{lemma}

\sp{\bf Proof.} 
Remark that there exists $\tau_1$ such that
$m_1=g^{\tau_1}(m)$. Then, by the commutation of flows we have
$g^{\cb}=g^{-\tau_1}\circ g^{\cb}\circ g^{\tau_1}$. Taking
the differential (with respect to the space variable) of this equality
at the point $m$ we get
$$ 
dg^{\cb}(m)=dg^{-\tau_1}(m_1)dg^{\cb}(
m_1)dg^{\tau_1}(m)\ , 
$$ 
which gives the relation between $dg^{\cb}(m)$  and $dg^{\cb}(m_1)$; 
the eigenvalues of the former are the Floquet multipliers at $m$, and
the eigenvalues of the latter are the Floquet multipliers at $m_1$. This 
shows that  $dg^{\cb}(m)$  and $dg^{\cb}(m_1)$  are conjugated and
therefore have the same eigenvalues. \EOP

\begin{lemma}
\label{impl}
Fix $m\in\Lambda$ and introduce adapted coordinates based at $m$,
consider the map $\Phi_\epsilon: (\beta,\tau,y) \to (\^\b_\epsilon ,
\^\tau_\epsilon ,  \^y_\epsilon )$; 
then there
exists $\beta_{*}$ and $\epsilon_*$ independent of $m$ such that 
$$
\^ y_\epsilon (\beta,\rho^{\epsilon}_m(\beta))=\rho^\epsilon_m(\beta)\ .
$$
defines a unique smooth map
$\rho^\epsilon_m(\beta)$ for all $\beta$, $\epsilon$ with
$|\beta-\beta_0|<\beta_{*}$ and $|\epsilon|<\epsilon_*$.
\end{lemma}

\sp {\bf Proof.}  In order to be guaranteed that we can solve the
equation $\^y_\epsilon(\beta,y)=y$ by means of the implicit function
theorem, we show that our assumption (iii) on the Floquet
multipliers (see theorem \ref{pert}) implies that $1$ is not an
eigenvalue of the Jacobian of the map $y\mapsto \hat y_0$ at
$(\beta_0,0,0)$.

To prove this fact remark that by lemma \ref{str} the Jacobian matrix
$J$ of $\Phi_0$ at $m$ takes the block form
\begin{equation}
\label{jac}
J=
\pmatrix{\displaystyle{
\frac{\partial\hat \beta}{\partial\beta}} &\displaystyle{ \frac{\partial\hat
\tau}{\partial\beta}} & \displaystyle{\frac{\partial\hat y}{\partial\beta}}
\cr
\displaystyle{\frac{\partial\hat \beta}{\partial \tau} }
&\displaystyle{ \frac{\partial\hat 
\tau}{\partial \tau}} &\displaystyle{ \frac{\partial\hat y}{\partial\tau}}
\cr
\displaystyle{\frac{\partial\hat \beta}{\partial y}} & \displaystyle{\frac{\partial\hat
\tau}{\partial y}} & \displaystyle{\frac{\partial\hat y}{\partial y}}
\cr}= 
\pmatrix{ 1 & \displaystyle{\frac{\partial
\bar \tau}{\partial\beta}} & \displaystyle{\frac{\partial
\hat y}{\partial\beta}} 
\cr
\null &\null &\null
\cr
0 & 1 & 0
\cr
\null&\null&\null
\cr
0 & \displaystyle{\frac{\partial
\bar\tau}{\partial y}} & \displaystyle{\frac{\partial
\hat y}{\partial y}}  
\cr}
\ .
\end{equation}
where we dropped the index $\epsilon$ which is here equal to zero.
It follows that the secular equation for $J$ takes the form (with $I$
the identity matrix)
$$ 
\det ( \lambda I -J ) \ = \ (\lambda-1)^{2s} \ \det \left(\lambda
I - \frac{\partial
\hat y}{\partial y}\right) \ = \ 0 
$$
which gives the relation between the Floquet multipliers $\lambda$ and
the eigenvalues of the Jacobian matrix of the map $y\mapsto \^y$.  As
$J$ has, by assumption, {\it exactly } $2s$ Floquet multipliers equal
to 1, it follows that 1 is not an eigenvalue of $(\partial \^y /
\partial y)$.  Note that this also shows that we have always a
multiplicity at least $2s$ for the eigenvalue 1

We can thus apply the implicit function theorem, which ensures
existence uniqueness and smoothness of the map $\rho^\epsilon_m$. We can choose
$\b_*$ and $\epsilon_*$
independent of $m$: indeed by compactness of $\La$ the $C^r$ norm of $\^y$
can be bounded uniformly with respect to $m$ and the eigenvalues of
$(\pa \^y / \pa y)$ 
are uniformly bounded away from 1, see lemma 2.9. \EOP

Define now the map $\s^\epsilon_\b : \La \to U$ by
$\sigma^\epsilon_\beta(m)\equiv (\beta,0,\rho^\epsilon_m(\beta))$
where we used adapted coordinates based at $m$.  Remark that since
$\sigma^\epsilon_\beta(m)\in\Sigma_m$, one has
$$
\sigma^\epsilon_\beta(m)\not=\sigma^\epsilon_\beta(m')\quad {\rm if
}\quad m\not=m'.
$$
Define
$$
\Lambda^\epsilon_\beta:=\sigma^\epsilon_\beta(\Lambda)\ ,
$$
and remark that this set is in one to one correspondence with $\Lambda$. We
are going to prove that actually $\Lambda^\epsilon_\beta$ is a smooth manifold,
and that the above correspondence is a diffeomorphism. To this end,
having fixed $m$, $\beta$ and $\epsilon$ with
$|\beta-\beta_0|<\beta_{*}$ $\left|\epsilon\right|<\epsilon_*$, define
$$
\M^{\beta,\epsilon}_m:=\left\{p\in\widetilde\V_m\ :\ p=(\beta,\tau,y)\ ,\
y=\rho^\epsilon_m(\beta) \right\}
$$
where we use adapted coordinates based at $m$. Then
$\sigma^\epsilon_\beta(m)\in\M^{\beta,\epsilon}_m$, and moreover
$\M^{\beta,\epsilon}_m$ is an $s$--dimensional smooth submanifold of
$M$.

\begin{lemma}
\label{minlambda}
One has $\M^{\beta,\epsilon}_m\subset\Lambda^\epsilon_\beta$.
\end{lemma}
\sp{\bf Proof.} In the proof we drop the index $\epsilon$ which is
assumed to be fixed and different from zero. Let
$p\in\M_m^\beta$, $p\not=\sigma_\beta(m)$, then
there exists a unique $m_1$ such that $p\in\Sigma_{m_1}$.  Consider
$\^p:=\Phi(p)$. Using adapted coordinates based at $m$, if
$p = (\b , \tau , y)$ then, by definition of $\M_m^\beta$,
the coordinates of $\hat p$ are $(\beta,\^\tau,y)$. From this
it follows $g^{-(\^\tau - \tau)} (\^p)=p$.
Introduce adapted coordinates $(\beta,\tau_1, y_1)$ based at the point
$m_1$; one has $p=(\beta,0,y_1)$, and denote
$$
\widehat p=\left(\beta, \widehat \tau_1, \widehat y_1\right)\ .
$$
Applying $g^{-(\^\tau - \tau )}$ we obtain
$$
 \left(\beta,\widehat\tau_1-(\widehat\tau - \tau), \widehat y_1\right)
= g^{-(\widehat\tau - \tau)} (\widehat
p ) = p =(\beta,0,y_1)
$$
from which in particular one has $y_1=\widehat y_1(\beta, y_1)$.  By
uniqueness of the solution to $y_1 = \widehat{y} (\b , y_1 )$, it
follows $y_1 = \rho_{m_1} (\b) $ and therefore $p=\sigma_\beta(m_1)$.
\EOP

\sp
{\bf Remark 6.}
The sets $\M^{\beta,\epsilon}_m$ are a covering of $\Lambda^\epsilon_\beta$: indeed
$\M^{\beta,\epsilon}_m$ contains at least $\sigma^\epsilon_\beta(m)$. 

\begin{lemma}
\label{smooth} $\Lambda^\epsilon_\beta$ is a smooth compact manifold, 
$C^r$-diffeomorphic to $\Lambda$.
\end{lemma}

\sp{\bf Proof.} Also in this proof we drop the index $\epsilon$. First
remark that the map
$$ \psi^\beta_m: \U_m \to \M_m^\beta \ \ , \ \ \psi_m^\beta
(\beta_0,\tau,0) = (\beta,\tau, \rho_m(\beta)) $$ (where we used
adapted coordinates based at $m$) is smooth from a neighbourhood
$\U_m$ of $m$ in $\Lambda$ to a neighbourhood of $\sigma_\beta(m)$ in
$\Lambda_\beta$. Use this map to introduce the $\tau$ as local
coordinates in $\Lambda_\beta$.  Then the maps $\psi^\beta_m$ yield a
$C^r$ atlas of $\Lambda_\beta$; the transition functions between the
coordinate system $\tau$ on $\M_{m}^\beta$ and the coordinate system
$\tau_1$ on $\M_{m_1}^\beta$ are given simply by the functions
$\tau_1(\beta,\tau,\rho_m(\beta))$, which are $C^r$-smooth. It follows
that $\Lambda_\beta$ is a $C^r$ manifold; it is also $C^r$
diffeomorphic (by the map $\sigma_\beta$) to $\Lambda$, and therefore
is compact.  
\EOP

\sp{\bf Remark 7.} By construction the fields $X^\epsilon_i$ are
tangent to each of the $\M^{\beta,\epsilon}_m$, and therefore the
manifold $\Lambda^\epsilon_\beta$ is invariant under the flows of all
the $X^\epsilon_i$. 

\begin{lemma}
\label{symp}
The $C^r$ manifold $\widetilde N_\epsilon$ obtained as the union of
$\Lambda^\epsilon_\beta$ for $\left|\beta-\beta_0\right|<\beta_*$ and
$\epsilon$ fixed is symplectic and fibered in isotropic tori
$\toro^s$.
\end{lemma}  

\sp {\bf Proof.} It is clear that $\widetilde N_\epsilon$ is $C^r$ and
that it is fibered in tori $\toro^s \equiv
\Lambda^\epsilon_\beta$. The tori $\Lambda^{\epsilon}_\beta$ are
integral manifolds of the Hamiltonian vector fields generated by the
functions $F^\epsilon_i$, and lie in common level manifolds of the
$F^\epsilon_i$; thus we can choose a basis of variables $\tau_i$ along
$\Lambda^\epsilon_\beta$ such that $\Om (\pa / \pa F^\epsilon_i , \pa
/ \pa \tau_j ) = \delta_{ij}$. This shows at once that the
$\Lambda^\epsilon_\beta$ are isotropic and that the restriction
$\Om_{\widetilde N_\epsilon}$ of the symplectic form $\Om$ defined on
$M$ to the submanifold $\widetilde N_\epsilon \ss M$ is
non-degenerate. As $\d \Om = 0$ implies $\d \Om_{\widetilde
N_\epsilon} = 0$, the proof is complete. \EOP

\begin{corollary}
\label{ciao}
There exist a $2s$ dimensional symplectic invariant submanifold $N_\epsilon$,
$\Lambda\subset N_\epsilon\subset\widetilde N_\epsilon$, and action angle
coordinates on $N_\epsilon$.
\end{corollary}
\sp{\bf Proof.}  The restriction of the system to $\widetilde
N_\epsilon$ is integrable in the Arnold Liouville sense. Thus the
standard construction of action angle coordinates for integrable
systems applies \cite{Arn1} and allows to construct action angle
coordinates in a neighbourhood $N_\epsilon$ of
$\Lambda^\epsilon_{\beta_0}$ in $\widetilde N_\epsilon$. \EOP

This concludes the proof of theorem \ref{pert}.

\section{Reducible tori}
\def\hot{{\tt h.o.t.}}
\def\Z{{\bf Z}}

The nondegeneracy condition of theorem \ref{pert} takes a particularly
useful form in the case of reducible tori that we are going to
discuss. 

Under the assumptions (i,ii) of theorem \ref{pert} assume also that
there exists a system of canonical coordinates $(I,\varphi,p,q)$, with
$\varphi\in\toro^s$ and $(I,p,q)$ defined in a subset of
$\Re^s\times\Re^{n-s}\times\Re^{n-s}$, in which the functions $F_i^0$
take the form
$$ 
F_i^0 \ = \ \sum_{j=1}^s \, \omega^{(i)}_j \, I_j \
+ \ \sum_{j=1}^r \, \nu^{(i)}_j \, \( {p_{j}^2+q_{j}^2 \over 2} \)+\hot
$$ 
where $\hot$ denotes higher order terms, i.e. terms which are at least
quadratic in $(I,p,q)$ if they depend on $I$, and terms which are
independent of $I$ and at least cubic in $p,q$, and $\Lambda$ is the
manifold $p=q=I=0$, and $r:=n-s$. { \it If such coordinates exist the torus
$\Lambda$ is said to be reducible.}

Concerning reducible tori we recall the following

\begin{proposition}
\label{kuk}
({\bf Kuksin}\cite{Kuk2}) Under the assumptions (i,ii) of theorem
\ref{pert}, let $\alpha^{(j)}$, $j=1,...,s$ be a basis of
$\pi_1(\Lambda)$; assume that for all $j$ all the Floquet multipliers
of the periodic orbits of the fields $X_{\alpha^{(j)}}$ have modulus
one, then $\Lambda$ is reducible.
\end{proposition}

We define, for ease of notation, matrices $A$ and $B$ with
elements given by 
$$ A_{ij} \ := \ \om^{(i)}_j \ \ ; \ \  B_{ij} = \nu^{(i)}_j \ . $$
Hence $A$ is a $(s \times s)$ matrix built with the frequencies of
motion in the invariant torus $\Lambda$, and $B$ is a $(s \times r)$
matrix built with the frequencies of small oscillations in the
transversal directions to the invariant torus.

\def\wt#1{\widetilde{#1}}

We will denote by $\Om (k;j)$ the matrix obtained from $A$ by
substituting its $k$-th column with the $j$-th column of $B$. For ease
of notation, we also write $P := (A^T)^{-1}$. We will also denote by
$|M|$ the determinant of a matrix $M$.

\begin{theorem}
\label{redu}
Let $X_\a$ be the vector field $X_\a = \sum_i c_i X_i$ having periodic
orbits $\gamma_\a$ on $\La$ with period one and in the homotopy class
$\a \in  \Z^s$.
Then: 

\noindent {\rm (1)} The condition that the multiplicity of the Floquet multiplier 1 of
$\gamma_\alpha$ does not exceed $2s$ is equivalent to $Q_j (\a)
\not\in {\bf Z}$, where
\begin{equation}
\label{complex}
Q_j (\a) \ = \ \sum_{i,k=1}^s \ B^T_{ji} P_{ik} \alpha_k \ .
\end{equation}

\noindent {\rm (2)}  Condition (1) is equivalent to 
\begin{equation}
\label{simple} 
\sum_{k=1}^s \ \alpha_k \, | \Om (k;j) | \ \not= \ m \ |A| \ \ \ \ 
\forall m \in {\bf Z} \ ,\quad \forall j=1,...,r\ .
\end{equation}
\end{theorem}

We stress that both the conditions given in the theorem
can be checked once we know the matrices $A$ and $B$; however
condition \ref{complex} requires to consider only one matrix $B^T
(A^T)^{-1}$ but also to perform the inversion of the $(s \times s)$
matrix $A$, while condition \ref{simple} requires to consider $s \cdot
(n-s)$ matrices $\Om (k,j)$, but does not require to consider any
inversion of matrices. Thus it can be more convenient  to use one or
the other of them depending on the problem at hand; however condition
\ref{simple} can always be explicitly checked and requires only
simple linear algebra. As far as we know, condition \ref{simple} has
not been considered before.

\sp{\bf Remark 8.} In the case $s=1$ condition \ref{simple} reduces to 
\begin{equation}
\label{lya}
\exists \alpha\in{\bf Z}\ :\ \alpha\nu_k\not=m\omega_1\ ,\
\forall m\in{\bf Z}\ ,\ \forall k=2,...,n \ , 
\end{equation}
which is easily seen to be equivalent to the standard condition under
which the Lyapunov center theorem holds, namely that
$\nu_k/\omega_1\not\in\ra $, $\forall k=2,...,n$. Indeed, if
$\nu_k/\omega_1\in\ra$ then, for any choice of $\alpha\in\ra$ one
has that $\alpha\nu_k/\omega_1\in\ra$, and therefore \ref{lya} is
also violated. Conversely, if $\nu_k/\omega_1\not\in\ra$ then
simply choose $\alpha=1$ and \ref{lya} also holds.  \EOR

\sp{\bf Proof of theorem \ref{redu}.} In this section we will drop
from $F^0_i$ the higher order terms which do not change the linearized
dynamics at the torus, we will drop also the index $\epsilon$ since
here we are only interested in the case $\epsilon=0$.

Fix $\alpha\in\pi_1(\Lambda)$;
first of all we determine the vector field $X_\alpha$, to this
end we remark that the projection $Z_c$ of a general vector field $X_c
= \sum_{i=1}^s c_i X_i$ to invariant tori is just  
$$ 
Z_c \ = \ \sum_{i=1}^s \ \sum_{k=1}^s \ c_i \, A_{ik} \pa_k\,\ , 
$$ 
where $\pa_k:=
\frac{\partial}{\partial\varphi_k} $. 
When we require that this
has closed orbits with period 1 and winding number $\alpha_i$ around
the cycles of $\La$, we are requiring
$$ 
Z_c \ = \ \sum_{k=1}^s \ (2 \pi)\, \alpha_k \, \pa_k \ . 
$$ 
Thus we get (recall $P := (A^T)^{-1}$)
$$  
c_i \ = \ \sum_{k=1}^s \ (2\pi) \, P_{ij} \, \alpha_j  , 
$$
which explicitly defines $X_{\alpha}:=\sum_{j=1}^s c_i X_i$.  

In order to compute the transversal Floquet multipliers we have to
study the dynamics of $X_\alpha$ in the transversal direction.  In
particular the flow generated by
$X_{\alpha}$ in the planes $p_j,q_j$ ($j=1,...,r)$ takes the form
$$ 
\pmatrix{p_j\cr q_j} \mapsto \pmatrix{p_j\cos(2 \pi \, Q_{j}(\a) \, t)
-q_j \sin(2 \pi \, Q_{j} (\a) \, t)\cr q_j \cos(2 \pi \, Q_{j} (\a) \,
t) + q_j\sin(2 \pi \, Q_{j} (\a) \, t)}\ ,
$$ 
This shows that the transversal Floquet exponents are indeed the $Q_j
(\a)$ and therefore condition (iii) of theorem \ref{pert} is
equivalent to $Q_j (\a) \not\in \Z$.

It remains to prove (2). Let us consider the matrix
$$ 
M \ := \ B^T \ P \ \equiv \ B^T (A^T)^{-1} 
$$
and the matrices $\Om (k,j)$ introduced above. We have to show that
the elements $m_{kj}$ of $M$ can be written as a ratio of
determinants, $m_{kj} = |\Om (k,j)|/|A|$.

Let us introduce a useful notation: given a matrix $M$ of
elements $m_{ij}$, we denote by $\wt{M}$ the matrix of its algebraic
complements. 
We recall that $\sum_j m_{ij} (\wt{m})_{kj} = \delta_{ik} |M|$, that
$\wt{(M^T)} = \(\wt{M}\)^T$, and that by Cramer's theorem $M^{-1}$ is
obtained as $(M^{-1})_{ij} = | M |^{-1} \(\wt{M}\)^T$.

For ease of writing, we will think of $k$ and $j$ as fixed and write
simply $R$ for $[\Om (k,j)]^T$.  One has $ |R| = \sum_\ell R_{i\ell}
(\wt{R})_{i\ell}$, for any $i=1,...,s$; we consider $i=k$ and compute
the determinant in this way. By definition, however, $R_{k\ell} =
(B^T)_{j\ell}$. As for $(\wt{R})_{k\ell}$, we note that $R$ differs
from $A^T$ only on the row $k$; thus the $k$-th rows of $\wt{R}$ and
of $\wt{A^T}$ are identical, i.e. $\wt{R}_{k\ell} = \wt{A^T}_{k\ell} =
\wt{A}_{\ell k}$.
 
In this way we obtain that 
$$ 
|R| \ = \ \sum_{\ell} \, R_{k \ell} \, (\wt{R})_{k \ell} \ = \
\sum_\ell \, (B^T)_{j \ell} \, \wt{A}_{\ell k} \ = \ \sum_\ell \,
(B^T)_{j \ell} \, [(A^T)^{-1}]_{\ell k} \, (|A|) \ . 
$$ 
Using $|R^T| =
|R|$ and writing again in full $R^T \equiv \Om (k,j)$, we have shown
that
$$ M_{jk} \ := \ (B^T P)_{jk} \ \equiv \ (B^T (A^T)^{-1})_{jk} \ = \ {|\Om (k,j) | \over |A|} \ , $$
hence that $ Q_j (\a) = \sum_k \a_k  (|\Om (k,j) | / |A|)$; 
this completes the proof of (2) and thus of the theorem. \EOP

\vfill\eject

\end{document}